 \documentclass[12pt,preprint]{aastex}

\usepackage {epsf}

\begin{document}



\def \etal { et~al.\ }         
\def \m {M$_\odot$}          
\def \mm {M_\odot}           
\def \degrees{$^\circ$}
\def \pderiv#1#2{\frac{\partial #1}{\partial #2}}
\def \deriv#1#2{\frac{d #1}{d #2}}

\def \degrees{$^\circ$}
\def \pderiv#1#2{\frac{\partial #1}{\partial #2}}
\def \deriv#1#2{\frac{d #1}{d #2}}

\title{{\bf The Convective Urca Process with Implicit Two-Dimensional Hydrodynamics 
}}

\author{J. Stein$^{1,3}$ and J.C. Wheeler$^{2,4}$}

\altaffiltext{1}{Racah Institute of Physics, Hebrew University of Jerusalem,
                Jerusalem, Israel}
\altaffiltext{2}{Department of Astronomy, University of Texas at Austin,
                 RLM 15.308, Austin, TX 78712-1083}
\altaffiltext{3}{E-mail: yossi@phys.huji.ac.il} 
\altaffiltext{4}{E-mail: wheel@astro.as.utexas.edu}

\bigskip
\begin{abstract}

Consideration of the role of the convective flux in the thermodymics of the
convective Urca neutrino loss process in degenerate, convective, quasi-static, 
carbon-burning cores shows that the convective Urca process slows down the 
convective current around the Urca-shell, but, unlike the ``thermal" Urca
process, does not reduce the entropy or temperature for a given convective 
volume. Here we demonstrate these effects with two-dimensional numerical 
hydrodynamical calculations.  These two-dimensional implicit hydrodynamics 
calculations invoke an artificial speeding up of the nuclear and weak rates. 
They should thus be regarded as indicative, but still qualitative. We find that, 
compared to a case with no Urca-active nuclei, the case with Urca effects
leads to a higher entropy in the convective core because the energy released 
by nuclear burning is confined to a smaller volume by the effective boundary 
at the Urca shell. All else being equal, this will tend to accelerate the 
progression to dynamical runaway.  We discuss the open issues regarding the 
impact of the convective Urca process on the evolution to the ``smoldering phase" 
and then to dynamical runaway. 

\end{abstract}

\keywords{Physical processes: convection - hydrodynamics 
- nuclear reactions - Stars: interiors - supernovae} 

\section{Introduction}

The nature of degenerate carbon burning prior to runaway in Type Ia supernovae
is a long-standing problem that has become more acute with the need
to understand the variation in the properties of the supernovae that
might affect their application to the measurement of cosmological 
parameters.  In this context, it is important to determine the effect of 
the convective Urca process because it may have direct observational implications 
for Type Ia supernovae. The convective Urca process can affect the density at 
which carbon undergoes dynamic runaway, subsequent electron capture and neutronizing 
reactions in the explosion, the size and gradient in the carbon/oxygen ratio at 
runaway, and the spectrum of convective motions and hence subsequent 
Rayleigh-Taylor unstable dynamic burning.  

The essence of the convective Urca process was first presented by
Paczy\'{n}ski (1972) whereby the convective circulation driven by
carbon burning in degenerate white dwarfs will yield first electron
capture and then $\beta$-decay of susceptible nuclei.  This yields
no net change in composition, but a loss of neutrinos (or antineutrinos)
at each step of the cycle along with their attendant energy.  
Paczy\'{n}ski argued that this cycle would cool the star and postpone 
dynamical runaway.  Previously, it had been pointed out by Bisnovatyi-Kogan 
\& Seidov (1970) that non-equilibrium weak interactions as involved in 
this process would yield local heating. Bruenn (1973) independently pointed out 
that each weak interaction added heat to the system, despite the loss of
the neutrino.  As electrons are captured below the Fermi sea,
another electron will drop from the Fermi surface to fill the ``hole,''  
resulting in heat. In the other portion of the cycle, $\beta$-decay will
deposit electrons with excess thermal energy above the Fermi sea. 
Paczy\'{n}ski (1974) acknowledged the arguments of Bisnovatyi-Kogan \& Seidov
and of Bruenn and pointed out that the energy flow between carbon burning
and Urca neutrinos depends on the interaction between the convective and
Urca processes. He noted that evolutionary calculations by Couch \& Arnett (1973)
and by Ergma \& Paczy\'{n}ski (1974) were done with the implicit assumption
that Urca neutrinos directly remove heat from the stellar interior. In this
paper Paczy\'{n}ski also noted that the convective Urca process will
lead to a gradient in the electron abundance. 

The physics of the convective Urca process was analyzed
by Barkat \& Wheeler (1990; hereafter BW), who argued that 
convective currents of composition play a critical role in the quasi-steady
state thermodynamics of the convective Urca process, so that cooling
terms associated with the convective flow exactly cancel the
microscopic heating terms described by Bruenn and lead to net
cooling. Their results seemed to be consistent with the results of the
careful numerical work of Iben (1978a,b, 1982). 

The conclusions of BW were called into question by Mochkovitch (1996).   
Mochkovitch presented a general thermodynamic analysis and concluded that 
the convective Urca process must heat the star. These issues were
addressed again by Stein, Barkat \& Wheeler (1999: hereafter SBW) who concluded, 
in agreement with Mochkovitch, that while the neutrinos associated with the
convective Urca process carry away energy, the entropy, and the temperature, 
cannot decline.  BW included a ``work" term that effectively removed energy from 
the total energy budget that could only have come from the kinetic energy, 
which must remain positive. The loss rate by the Urca process cannot exceed 
the rate of generation of kinetic energy.  Rather, the convective Urca neutrino 
losses slow the convective currents (SBW, Bisnovatyi-Kogan 2001; Lesaffre, Podsiadlowsi
\& Tout 2005).

Here we present calculations demonstrating the role of the convective Urca process 
in two-dimensional simulations and illustrate that the conclusions of SBW were correct. 
In particular, we examine the role of convective currents, buoyancy and mixing in 
creating and limiting the kinetic energy. 

We present the outline of the input physics in \S2 and the numerical technique in \S3.  The 
results are presented in \S4 and conclusions are presented in \S5.


\section{Description of Physics and Models}

\subsection {Initial setting}

We constructed a Carbon-Oxygen core ($1.38 M\odot$) with equal abundance of C and O
by mass and with $\rho_c = 3.5\times10^9$ g cm$^{-3}$ and $T_c = 3\times10^8$ K
in which the inner region ($0.79 M_{\odot}$) had constant entropy, and the outer 
region was isothermal. For the $^{23}$Ne/$^{23}$Na Urca pair, the Urca shell is at
a density of $1.7\times10^9$ g cm$^{-3}$. In these calculations, the Urca shell 
falls in the middle of the constant-entropy region, and divided that region into 
an inner zone ($0.31 M_{\odot}$) and an outer zone. This setup was necessary to 
distinguish the effects of the Urca process on the convection from the effect of 
the steep entropy gradient in the isothermal part. The inner zone contained a small 
concentration ($0.00004 - 0.0004 \%$) of ``mother'' Urca nuclei ($^{23}$Ne) and 
the outer zone with the rest of the star contained the same concentration of 
``daughter'' Urca nuclei ($^{23}$Na ). This division between the inner and outer 
zones is stable against convection, even without the Urca process,  as long as the 
entropy of the outer zone is equal to (or even very slightly less than) the 
entropy of the inner zone.  In order to see the stabilizing effect of the Urca process 
on the convection we had to create a marginally stable configuration. We achieved this 
by slightly reducing the entropy of the outer zone, and creating a slightly 
{\it unstable} configuration. This instability triggered a violent convection, 
which settled down after a short time, leaving the simulation in the desired configuration.

\subsection {Reaction rates}
For the Carbon burning we used a simplified reaction rate which turns Carbon into Oxygen:
\begin{equation}
\frac{dX_c}{dt} = -0.5775 X_C^2 (\rho_9)^{2.79} (T_9)^{22} 
\end{equation}
where $\rho_9$ is the density in units of $10^9$ g cm$^{-3}$ and $T_9$ is the temperature
in units of $10^9$ K. The q-value for the reaction is  $q = 2.857\times10^{17}$ erg gm$^{-1}$.
The Urca data for density thresholds and rates were taken from Ergma and Paczy\'nski (1974).
In order to distinguish between the convective Urca process and the cooling thermal
Urca process, we neglect the thermal Urca process. 
Other neutrino losses were taken from Itoh and Kohyama (1983). In our simulations 
we multiplied all the rates by factors $2000$ or $20000$ in order to speed up the 
process. Even so, the simulations required hundreds of thousands time-steps to produce 
$\sim 10^4$ seconds of simulation ``star time."

\section{Numerical Procedure}

The goal of this program is to simulate convection in a marginally stable
 carbon-oxygen core, where the sound-speed is high and the material velocity
 is low.  Two-dimensional simulation of subsonic convection in the central part 
of a white-dwarf presents some major stability problems: \newline
1. Having nearly the Chandrasekar mass,
   the star is only marginally stable to small perturbations.\newline
2. Convection occurs in a region with huge density and temperature gradients,
   but almost constant entropy. The convection is driven by the
   tiny entropy gradients. Small density or temperature fluctuations caused by 
   numerical inaccuracies may create large entropy gradients, destabilizing the convection. \newline
3. The velocities are subsonic, so $\nabla \cdot(\rho {\mathbf u})$ is almost zero,
   in spite of the large density gradients.

These issues consequently require a very stable numerical procedure.
In order to address these stability issues, we therefore incorporated the following 
features into our program entitled DWARF:\newline
1. We use an implicit Eulerian scheme with second-order donor. \newline
2. In order to avoid the ``chessboard black-white'' decoupling of the
   cells in a usual staggered mesh, where density, energy and pressure
   are defined inside the cells and the velocity is defined on the corners, we define the
   velocities on the edge-centers of the 2D cells. \newline
3. Since the pressure differences between adjacent cells due to density
   differences may be larger by several orders of magnitude than the pressure
   difference due to the entropy differences, we take special care
   that the advection terms will not create spurious large entropy gradients.

\subsection {Hydrodynamic equations}
The hydrodynamic equations are:

\noindent
Acceleration:
\begin{equation}
   \pderiv{{\bf u}} {t} =
        -\frac{1}{\rho}\nabla{p} - ({\bf u} \cdot \nabla){\bf u} -{\bf G}
\end{equation}
Mass conservation:
\begin{equation}
   \frac{\partial{\rho}} {\partial{t}} = 
        -{\bf\nabla}\cdot(\rho{\bf u})
\end{equation}
Internal energy:
\begin{equation}
   \frac{\partial{e}} {\partial{t}} = 
        -{\bf\nabla}\cdot(e{\bf u}) -p{\bf\nabla}\cdot{\bf u} +q
\end{equation}
\noindent
where ${\bf u}$ is the velocity, $\rho$ is the density, $e$ is the energy per
unit mass, ${\bf G}$ is the gravitational force and $q$ is the energy source
per unit mass per unit time. We use spherical coordinates, $r,\theta,\phi$, and 
assume that the gravity has spherical symmetry. All other quantities are treated in 
cylindrical symmetry.

In these coordinates and under these assumptions our equations become:

Acceleration:
\begin{equation}
   \pderiv {u_r} {t} =
        -\frac{1}{\rho}\pderiv{p}{r} -g_r - u_r \pderiv {u_r}{r}
        -\frac{u_\theta}{r} \pderiv{u_r}{\theta}  +\frac{1}{r}u_\theta^2
\end{equation}
\begin{equation}
   \pderiv {u_\theta} {t} =
        -\frac{1}{\rho r}\pderiv{p}{\theta}
        -\frac{u_\theta}{r} \pderiv{u_\theta}{\theta}
        -u_r \pderiv {u_\theta}{r} -\frac{1}{r}u_r u_\theta
\end{equation}

Mass:
\begin{equation}
  \pderiv{\rho}{t} = -\frac{1}{r^2}\pderiv{(r^2\rho u_r)}{r} 
    -\frac{1}{r \sin{\theta}}
    \pderiv{(\rho u_\theta \sin\theta)}{\theta}
\end{equation}

Energy:
\begin{equation}
  \pderiv{\rho e}{t} = -\frac{1}{r^2}\pderiv{(r^2\rho e u_r)}{r} 
    -\frac{1}{r \sin{\theta}}
    \pderiv{(\rho e u_\theta \sin\theta)}{\theta}
\end{equation}

We set rigid boundaries in angle at $\theta = 45^o$ and at $135^o$ and in 
mass at an inner boundary of $10^{-4}$ \m\ and an outer boundary of 
$1.38$ \m. The two angle and inner mass boundaries were chosen to 
avoid the singularity of spherical coordinates. The outer mass boundary
was adopted for stability reasons. Despite the latter rigid outer boundary,
the radial relaxation (expansion) of the convective region was not affected.

\subsection {Features of the Program DWARF}

The DWARF code incorporates a rectangular grid in polar coordinates. It uses
a staggered-mesh with implicit-pressure scheme, in which only the pressure is 
treated implicitly (See Livne 1993 and references therein), is adequate for subsonic 
flow in an otherwise stable medium. In our particular case, this model
shows strong  "Chessboard black-white" decoupling of the cells, causing diagonal 
instability. This decoupling is caused by the fact that in this model 
$\pderiv{P_{i,j}}{P_{k,l}}$ is nearly zero, when
 $(k,l)=(i,j\pm 1)$ or $(k,l)=(i\pm 1,j)$.
To avoid this instability we chose a different staggered-mesh, in which the velocities and 
accelerations are defined at edge-centers. Other physical quantities are defined at the
cell-centers. For the implicit-pressure scheme, (see Livne 1993 and references therein), 
only the $M_{ij,kl} = \delta_{i,k}\delta_{j,l} - \pderiv{P_{i,j}}{P_{k,l}}$ matrix is 
inverted; however, this derivative accounts for both acceleration and advection.

To compute the acceleration, we divide the acceleration into two parts:\newline
 1- Direct accelerations,
 $\frac{1}{\rho}\pderiv{P}{r} -g_r - u_r \pderiv {u_r}{r}$ and
 $ \frac{1}{r\rho}\pderiv{P}{\theta} -\frac{u_\theta}{r} \pderiv{u_\theta}{\theta}$,
 are treated implicitly.\newline
 2- Centrifugal ($\frac{1}{r}u_\theta^2$) and Coriolis
 ($-\frac{1}{r}u_r u_\theta$) accelerations are treated explicitly,
using the values at the end of the time-step.

Stability requires that the advection be computed by a donor scheme;
however, a donor scheme is difficult to treat implicitly, because the dependence of the
advection on the velocity has discontinuous derivatives.
Therefore we divided the advection into two parts:\newline
1- Using centered values (no donor) in the implicit part.\newline
2- Correction to a second order donor scheme is done explicitly.\newline
Choosing the ``centered'' donor can be done in several ways. Since the
convection is driven by entropy gradients, we defined ``centered'' to
be average density, entropy and composition of the two adjacent cells.
The energy and temperature were calculated from these values by using the equation of state.

\section{Results}
We present the results of 5 simulations:\footnote{computational models HN, FH, GG, 
EG and FI, respectively} 
\newline
S-4-2E4-0: $(X_{Ne}=4\times 10^{-4})$, reaction-rates multiplied by 20000, Urca-rate = 0.\newline
S-4-2E4: $(X_{Ne}=4\times 10^{-4})$, reaction-rates multiplied by 20000,
    Urca-rate multiplied by 2000.\newline
S-1-2E4: $(X_{Ne}=1\times 10^{-4})$, reaction-rates multiplied by 20000,
    Urca-rate multiplied by 2000.\newline
S-0.4-2E4: $(X_{Ne}=4\times 10^{-5})$, reaction-rates multiplied by 20000,
    Urca-rate multiplied by 2000.\newline
S-4-2E3: $(X_{Ne}=4\times 10^{-4})$, reaction-rates multiplied by 2000,
    Urca-rate multiplied by 2000.\newline
When we multiplied the Urca-rate by another factor of 10 (to 20,000), the results 
did not change significantly, although the Urca-gradients became steeper, spreading 
over fewer cells. We thus left Urca rate multiplication factor at 2000 for
all models where Urca effects were incorporated.

All simulations were computed a long time after the convection settled down.
Simulations S-4-2E4, S-1-2E4 and S-0.4-2E4 (varying the Urca abundance) ran up to 
6000 simulation-time seconds. Simulation S-4-2E4-0 (with no Urca) ran up to 10,000 
simulation-time seconds and simulation S-4-2E3 (with reduced nuclear rate acceleration), 
which was slower than the rest, ran up to 20,000 simulation-time seconds.  The 
velocity scale in all the 2D figures (3 - 6 below) is normalized so that a
single horizontal tic mark corresponds to a velocity of 10 km s$^{-1}$. For
the online color figures the velocity is as follows: black arrows: 
below $10^5$ cm s$^{-1}$; red arrows: $10^5$ - $2\times10^5$ cm s$^{-1}$; 
green arrows: $2\times 10^5$ - $4\times10^5$ cm s$^{-1}$;
blue  arrows: $4\times 10^5$ - $8\times10^5$ cm s$^{-1}$.

The energy input by nuclear reactions is converted in part to the kinetic energy
of the convective circulation. The fraction of the input energy that ends up in 
kinetic energy depends on the dissipation to Urca ``viscosity," to turbulent
viscosity and, for simulations, into numerical viscosity. For reference, the
total energy input rate for simulation S-4-2E4 is $2\times10^{42}$ erg s$^{-1}$.
The Urca neutrino loss rate is about $10^{41}$ erg s$^{-1}$. The total
Urca dissipation rate including heating is about $1.3\times10^{41}$ erg s$^{-1}$,
about 6.5 percent of the energy input. As shown in SBW, the rate of
conversion of thermal energy to kinetic energy up to the Urca shell
($T \sim 1.9\times10^8$ K) is less than $(T_c -T_U)/T_c \sim 37$ percent,
where $T_c$ is the central temperature and $T_U$ is the temperature
at the Urca shell. This implies that the Urca process as computed here
makes a significant contribution to the kinetic energy dissipation
in terms of the estimated maximum dissipation. 

Figure 1 shows the kinetic energy of all the simulations.  In simulations 
S-4-2E4, S-1-2E4, and S-0.4-2E4 in which the nuclear rates were all multiplied by 
20,000, but the abundance of Urca nuclei varied by a factor of 10, the kinetic 
energy associated with the convection settled rapidly into a quasi-steady state after 
recovering from the initial instability.  In simulation S-4-2E4-0, which had no 
Urca effect, the convection spread to fill the whole inner, constant entropy region, 
and so the approach to steady state took longer. Simulation S-4-2E3, with the 
nuclear rate enhanced by only a factor of 2000, took much longer to reach steady state. 
Simulations S-4-2E4, S-1-2E4, and S-0.4-2E4 converged to more or less the same kinetic 
energy, showing that in these simulations the kinetic energy does not depend sensitively 
on the abundance of the Urca nuclei, as long as the Urca physics is active.
Simulation S-4-2E4-0, which lost no kinetic energy to the Urca effect, reached a 
somewhat higher value of kinetic energy. Simulation S-4-2E3, with a nuclear rate 
10 times less asymptoted to a value around $4$ times smaller than simulations 
S-4-2E4, S-1-2E4, and S-0.4-2E4. This gives some notion of the dependence of the 
convective kinetic energy on the rate of nuclear energy input.

Figure 2 shows the total Urca neutrino loss rate of all the simulations.
Although there are large fluctuations, the general trend is that a higher abundance 
of Urca nuclei leads to {\it smaller} neutrino losses. Specifically, simulation S-4-2E4, 
which had the highest concentration of Urca atoms, lost {\it less} energy to Urca 
neutrinos by a factor of 2 to 3 than simulations S-1-2E4 and S-0.4-2E4. This is 
presumably related to the fact that the higher $^{23}$Ne abundance restricted the 
extension of the convection beyond the Urca shell and the attendant mixing outward of 
carbon-burning products compared to simulations with small abundances of Urca-active 
nuclei (see Figures 3 below). On the other hand, we cannot rule out that this
difference is affected by our numerical resolution.  Simulation S-4-2E3, which had 
the smallest nuclear rate enhancement lost less energy to the Urca neutrinos than 
simulation S-4-2E4, by a factor of about $5$, which is very similar to the ratio of 
a factor of 4 between the asymptotic kinetic energies of the two simulations.  This 
suggests that the Urca neutrino loss rates are roughly proportional to the kinetic 
energy in the regime where Urca losses control the extent of convection (see
the Appendix for a discussion of conditions where strong convection may overwhelm 
the Urca process). 

The question this work was designed to address is: ``What does the Urca effect do?''
The answer to this question, our main result, is presented in Figure 3.
These figures show the percentage of burned carbon (more accurately: $0.5-X_{C}$),
for simulations S-4-2E4-0, S-4-2E4, S-1-2E4, and S-0.4-2E4 respectively.
The carbon burns near the center. When no Urca effect is present - the convective currents
spread the carbon-burning products evenly over the whole inner region (Figure 3A, 
simulation S-4-2E4-0).  In Figure 3B, simulation S-4-2E4, only a small percentage 
of the carbon-burning products sneaks through the Urca-shell.  This percentage 
increased as $X_{Ne}$ was decreased (Figures 3C and 3D).

Figure 4 shows the entropy in simulations S-4-2E4-0 and S-4-2E4, respectively. 
Note the decrease of the entropy beyond the Urca-shell in simulation S-4-2E4, 
which compensates for the increase in the number of electrons per atom.

Figure 5 shows the abundance of Urca-active Ne$^{23}$ atoms in the inner 
region of simulations S-4-2E4-0 and S-4-2E4, respectively.  As expected, 
in simulation S-4-2E4-0 the abundance is homogeneous, while in simulation 
S-4-2E4, the Ne$^{23}$ abundance is 0.0004 inside the inner zone within the 
Urca shell, and zero in the outer zone beyond the Urca shell, except for a 
narrow transition zone near the Urca-shell.  Figure 6 shows the Urca neutrino 
loss-rate in simulation S-4-2E4. The whole Urca activity occurs near the Urca-shell, 
with the exception of some boundary-related activity. In all the 2D figures, the 
overshootings at the left and right boundaries are numerical artifacts.  

\section{Conclusions}

These two-dimensional calculations make no assumptions about the convective 
Urca process {\it per se}; they employ the relevant weak interaction microphysics 
and then follow the affects of the convective flow with no extra assumptions about
thermal or chemical equilibrium. Work done by convection and associated neutrino 
losses are computed straightforwardly. Because the computations have been done 
with artificially accelerated thermonuclear and weak interaction rates, they 
cannot be taken literally; some processes like turbulent dissipation will not 
scale linearly with these rates.  It would, of course, be desirable to do 
these simulations in three, rather than two dimensions. We expect the 
qualitative results to be the same, but this must eventually be checked. 
Our results should therefore be judged for their qualitative value.  

With these caveats, we conclude here: a) In agreement with Mochkovitch (1996) and 
SBW and in disagreement with the earlier work in BW, that the convective Urca 
process cannot result in a net decrease in entropy, and hence in temperature, 
for a constant or increasing density; b) that the Urca process controls the 
convective flow and affects the kinetic energy even as the convective motions drive 
the Urca process; c) that the convective Urca process will, at least temporarily, 
limit the expansion of the convective zone beyond the Urca shell.  Bisnovatyi-Kogan 
(2001) also noted that the convective Urca process must slow the speed of convection.  
Despite their qualitative nature, we feel these simulations have established that 
{\it the convective Urca process is fed by the positive definite kinetic energy and 
cannot cool the star, but can and will affect the nature of the convective flow}.

Over the last couple of decades most work on modeling of the progenitors of 
Type Ia supernovae has simply ignored the effects of the convective Urca process. 
It remains to be seen whether the convective Urca process will have a significant
effect on the ultimate dynamical runaway.  The convective Urca process by itself 
will not lead to dynamical runaway at higher density because it cools the star, 
as hypothesized in some earlier work. The convective Urca process may shorten the 
``smoldering" phase (H\"oflich \& Stein 2002) between carbon ignition, defined 
as the epoch when carbon burning exceeds neutrino losses, and dynamic runaway. 
Compared to a model without the effect of the convective Urca process, the 
convective core, and hence the heat liberated by the carbon burning, is confined 
to a smaller volume. All else being equal, a model that invokes the convective 
Urca process should proceed to dynamical runaway more quickly than one without 
the convective Urca process. The net result might be to indirectly lead to 
dynamic runaway at lower density with effects on the explosion dynamics, 
nucleosynthesis, and light curve.

These issues will depend sensitively on whether there is an
active Urca shell and that, in turn, depends on the density
of carbon ignition.  While that issue was once actively explored,
it has received little commentary in recent literature on the
topic of degenerate carbon burning.  This is in part due to
advances in understanding of weak interaction rates (see Brachwitz
et al. 2000, and references therein) and partly due to settling
on a standard set of assumptions for the treatment of effects such
as plasmon neutrino losses and strong screening effects 
(that may or may not be correct). 

Calculations of non-rotating accreting white dwarf evolution tend to
give carbon ignition at densities exceeding $4\times10^{9}$ gm cm$^{-3}$
for modest accretion rates, $\sim 10^{-9}$ \m yr$^{-1}$, depending on the
mass of the original white dwarf and the temperature at the beginning of
the accretion phase (Bravo et al. 1996; H\"oflich, Nomoto, Umeda
\& Wheeler 2000). Lesaffre et al. (2006) estimate the distribution
of central densities for single-degenerate progenitors and find
a broad distribution with a minimum ignition density of $\sim
2\times 10^9$ g cm$^{-3}$, above the threshold density for the
Ne/Na Urca pair. Such conditions will inevitably 
be accompanied by one or more convective Urca shells. These conditions
might be expected to lead to nova outbursts that prevent growth to
runaway. If growth of the white dwarf continues, the result could be
accretion-induced collapse (Saio \& Nomoto 1998).  It will be interesting 
to investigate whether or not confinement of convection to Urca shells and 
the subsequent tendency to heat a smaller volume more rapidly would affect 
this conclusion.  Only for rather high accretion rates, approaching 
$10^{-6}$ \m yr$^{-1}$, will the central density at carbon ignition be less 
than the threshold density for the Ne/Na convective Urca cycle so that
the convective Urca process might not occur at all. For these high rates,
there is some danger of excessively polluting the circumstellar 
environment with a hydrogen-rich wind.  

Observational constraints on Type Ia supernovae tend to put the central density 
at ignition in a range where the Urca process might occur, independent of how 
the star (or model) got there.  
Plausible calculations and their dynamical outcomes that seem to reproduce 
observations of Type Ia supernovae (H\"oflich, Nomoto, Umeda \& Wheeler 2000;
Dominguez, H\"oflich \& Straniero 2001) give ignition densities of about 
$1.5 - 2.5\times10^9$ gm cm$^{-3}$, right where the Ne/Na convective Urca 
``on" switch might lie. This means that whether a model (and hence a real star, 
in principle) does or does not trigger the convective Urca process with its 
potential effect on convective velocity fields, extent of convection, amount
of carbon burning between ignition and runaway, and the C/O gradient at runaway, 
may depend sensitively on such recently neglected issues as electron screening and 
plasmon neutrino loss rates.  We also note that the white dwarfs that explode as
Type Ia supernovae will almost surely be rotating. The rotational state will 
also affect the carbon ignition density (Piersanti et al.  2003a,b) and hence 
the presence of convective Urca effects. 

Another important issue is the abundance of a given Urca-active pair. We have 
shown that the kinetic energy of convective motion is rather independent of the 
abundance of the Urca nuclei, but that the mixing of carbon and its burning
products does depend on the Urca abudance. We also find that neutrino losses are 
reduced (within the range of parameters explored here) for higher abundance of 
Urca nuclei, although we cannot rule out that this is an effect of our 
numerical resolution. The abundance of Urca-active nuclei are a function 
explosive nucleosysthesis in prior generations of stars and so the role of 
the convective Urca process might be a function of redshift. Another possible 
effect on abundances is evolution in the core. If burning prior to runaway
should produce more Urca nuclei, then the Urca process could be enhanced. 

Other interesting issues are related to the location of the Urca shell (or shells) 
and the speed of convection. How does the Urca process work if the central density 
is only slightly above the Urca threshold so that the Urca shell is quite small, 
smaller, perhaps, than a convective scale length? In general, how fast is the 
convection? Sufficiently rapid convection will sweep mothers back down to regions 
where they are stable before they can decay. We note that when carbon ignition first 
occurs, the convection is likely to start slowly, giving the Urca process ample 
time to work, as long as the ignition density exceeds the Urca threshold. 

The key question is, even if there is an active convective Urca stage
in the evolution toward dynamic runaway, does this make any substantial
difference in the nature of that runaway and the subsequent dynamics?
If the Urca process limits the radial extent of the convection, somewhat
more carbon might be consumed within the Urca shell than beyond. On the
other hand, the total amount of carbon consumed is likely to be small and
if the limitation on the extent of convection breaks down before dynamic
runaway, then there will be a homognenization process as the convection 
sweeps outward and dredges carbon inward. It seems unlikely that an effect of 
the convective Urca process on the abundance, {\it per se}, will be significant.

A potentially more interesting effect is on the velocity structure of 
the convective motion.  H\"oflich \& Stein (2002) investigated the
final 3 hours before dynamic runaway with implicit two-dimensional
calculations. They showed that the ``smoldering" phase of nuclear burning 
that immediately preceeds dynamical runaway can produce convective motions 
with velocities of order 40 - 120 km s$^{-1}$. There were two implications of
this work. One was that dynamical runaway was triggered at a single region
near to, but definately away from, the center of the star by compression
driven by converging convective plumes. The second implication was that
the convective speeds were comparable to, or even exceeded, the velocity 
of the burning front in the early, slow, phase of Rayleigh-Taylor unstable 
turbulent deflagration.  This means that the convective velocity field 
can substantially ``prepare" the conditions prior to dynamical runaway,
can trigger the dynamical runaway, and that in the early phases of 
dynamical runaway it is still the convective velocities, not the burning 
front velocity, that determine the speed of propagation of the burning front.  
Recent calculations by Kuhlen et al. (2005) have addressed aspects of
the late smoldering phase in three-dimensional calculations. They have
not followed the burning completely to dynamic runaway, but find qualitatively 
similar results over the limited time frame they investigated: convective 
flows of order 50 - 130 km s$^{-1}$ and hot spots near the center where 
dynamic runaway might occur. The outcome of dynamical ignition may depend 
sensitively on the number and location of the runaway spots leading in some 
cases to single plumes of burning and in others to more homogeneous outward 
propagation of the early deflagration burning (Livne, Asida \& H\"oflich 2005).

The convective Urca process could thus affect the nature of the convective 
velocity field and hence the smoldering phase, the location of dynamic
runaway, and the early stages of dynamical runaway.  The current calculations 
allow us to raise such issues, but because of the artificial speed up of the 
nuclear and weak interaction rates, we cannot address them directly here. The
speed up of the nuclear rates by a factor of 20,000 will result in a speedup 
of the convective velocities.  In the current models, the convective motions are 
of order 10 km s$^{-1}$. The question is what happens in a realistic model.
Will the presence of the convective Urca shell continue to confine the 
convection and will the convective velocities in the smoldering phase that 
precedes dynamical runaway be restricted and hence less able to affect the 
early stages of deflagration? Or will the convection speed up and somehow
``break out" of the Urca shell yielding results similar to those of 
H\"oflich \& Stein (2002) who neglected the convective Urca process.

In this context, the important issue is whether the convection becomes strong 
enough within an active Urca zone to overwhelm the Urca control and thus
spread the convection throughout its ``natural" regime as if there
were no Urca processes. The convective Urca process might control
the extent of the convection through much of the smoldering phase
only to be overwhelmed in the final few hours or minutes before
dynamic runaway; sufficient time to establish a convective structure that
lost all memory of the convective Urca phase. 

That this breakdown of the convective Urca process could occur
can be seen qualitatively by considering that for sufficiently 
strong convection, convective flows may push substantially
beyond the Urca shell to the point where all the mothers have
decayed to daughters and there will be no more neutrino losses
regardless of how far the convection thus extends. We have shown
in Figures 1 and 2 that the convective kinetic energy and neutrino
loss rates increases with the nuclear reaction rates, but, at least 
for these simulations, at a rate that is slower than linear.
These simulations also suggest that within the range we have
explored, the neutrino loss rates scale roughly linearly with
the convective energy. These conclusions hold as long as the
Urca shell can control the convection. We sketch in the Appendix
an argument that suggests that sufficiently strong convection
can cause the convection to ``break out" and overwhelm the
limits of the convective Urca process. In principle, our use
of accelerated nuclear reaction rates might correspond to 
a later phase in the smoldering process, closer to dynamical
runaway, although it is not clear this is true. Based on our
simulations, we estimate in the Appendix that convective velocities 
substantially in excess of 20 to 70 km s$^{-1}$, depending on the
abundance of the Urca-active nuclei, are required for ``breakout."
Interestingly, this is similar to the velocity range found in the
studies of H\"oflich \& Stein (2002) and of Kuhlen et al. (2005)
for the late smoldering phase. This suggests that breakout
might, or might not, occur prior to dynamical runaway. We conclude
that this issue is open and in need of thorough future study.
While there is no obvious reason to think the convective Urca
process will affect subsequent dynamic phases, we currently
have no grounds to rule out that it may do so.  

Although the convective Urca process has been long neglected, the
current calculations suggest that it needs to be in the
suite of actively-considered physical phenomena as we attempt
to achieve a refined level of understanding of Type Ia supernovae.
We have identified a number of important issues regarding the
convective Urca process that need to be explored in multidimensional 
calculations that investigate the evolution from carbon ignition 
to dynamical runaway. A key issue is the behavior of the convective
Urca process without the artifical enhancement in rates we have
employed and in three dimensions.

We are grateful to Peter H\"oflich for discussion of related topics
of degenerate convective carbon burning. This research was supported in part 
by NSF Grant AST-0098644,

\appendix

\section {An Estimate of Urca Stopping Power}

In order to get a rough estimate of the stopping power of the Urca effect, 
let us assume that one blob of unit mass of material containing  
concentration of mothers, $X_m$, crosses the Urca-shell outward. its velocity 
{\bf u} is perpendicular to the Urca-shell. Let us further assume that 
the Urca effect is local, i.e., when mothers become daughters in the blob, 
the blob loses kinetic energy equivalent to the neutrino + heat energy 
released in the process. This estimate would be straight forward
except that the energy lost depends on the location in which it 
was released: the further from the Urca-shell, the more energy is lost
to neutrinos plus $\beta$-decay heating.

Let $x$ be the distance of the blob from the Urca-shell, then, under the 
above assumptions, the motion of the blob is governed by the following 
equations: 
\begin{equation}
  \deriv{x}{t} = {\bf u}
\end{equation}
\begin{equation}
  \deriv{X_m}{t} = -C_{\nu} \eta^3 x^3 X_m
\end{equation}
where $ C_{\nu} \eta^3 x^3$ is the rate of capture per atom with $C_{\nu}$ being
a numerical coefficient (units erg$^{-3}$ s$^{-1}$) characteristic of a 
given Urca pair, $\eta = | \deriv{E_F}{x} |$, at $x=0$, $N_A$ is 
Avogadro's number and $A$ is the atomic number of the Urca atoms.

If the blob is not stopped by the Urca process, but loses a fraction of 
its kinetic energy, then $X_m$ goes to $0$ for large $x$ where all
$\beta$-decay has gone to completion. Then 
$| \deriv{X_m}{t} |$ has a maximum at some point $x_c$. Let us 
assume that the whole reaction occurs near that point.  At the maximum 
we have: $x_c=(\frac{3u}{C_{\nu} \eta^3})^\frac{1}{4}$. The energy released 
at the maximum is:
\begin{equation}
E_c = \frac{N_A}{A} X_m \eta x_c = 
\frac{N_A}{A} X_m \eta \times (\frac{3u}{C_{\nu} \eta^3})^\frac{1}{4}
=\frac{N_A}{A} X_m (\frac{3u \eta}{C_{\nu}})^\frac{1}{4}.
\end{equation}
Therefore, in order to continue moving outward, the blob should have 
kinetic energy significantly larger than $E_c$, or 
\begin{equation}
u > (\frac{N_A}{A} X_m)^\frac{1}{2} (\frac{48u\eta}{C_{\nu}})^\frac{1}{8}
\end{equation} 
or
\begin{equation}
u > (\frac{48\eta}{C_{\nu}} (\frac{N_A}{A} X_m)^4)^\frac{1}{7}.
\end{equation} 
Note that this criterion depends only weakly on the Urca rate and
is only mildly sensitive to the Urca abudance.

In our case $\nu = 2000\times 3.3\times 10^{13} = 6.6\times 10^{16}$ 
erg$^{-3}$ s$^{-1}$ (Ergma \& Paczy\'nski 1974), $\eta = 0.86\times 10^{-13}$, 
and $A = 23$, so when $X_m = 4\times 10^{-4}$ then $u$ must be significantly 
greater than 70 km s$^{-1}$.  When $X_m = 4\times 10^{-5}$, $u$ must be 
significantly greater than 19 km s$^{-1}$.

\clearpage

\begin{figure}
\centering
\figurenum{1}
\epsscale{0.8}
\includegraphics[angle = 270, width = 6 in]{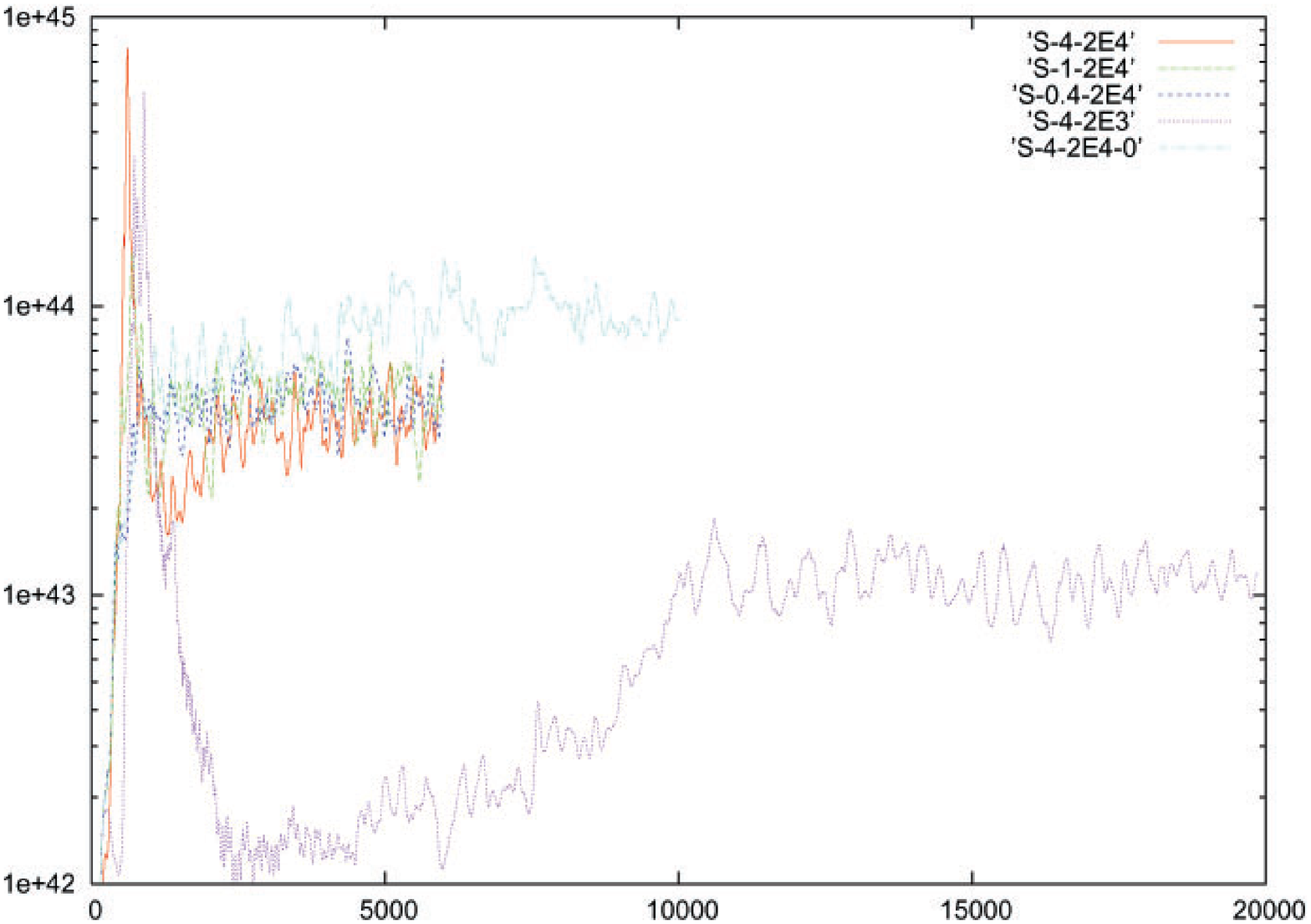}
\caption{Plot of convective kinetic energy versus simulation time, 
for simulations S-4-2E4, S-1-2E4, S-0.4-2E4, S-4-2E3, and S-4-2E4-0. 
Simulations  S-4-2E4, S-1-2E4, S-0.4-2E4, and S-4-2E4-0 had 
the nuclear rates multiplied by $2\times 10^4$ and simulation S-4-2E3, had the nuclear
rate multiplied by $2\times 10^3$.  
 }
\end{figure}

\clearpage

\begin{figure}
\centering
\figurenum{2}
\epsscale{0.8}
\includegraphics[angle = 270, width = 6 in]{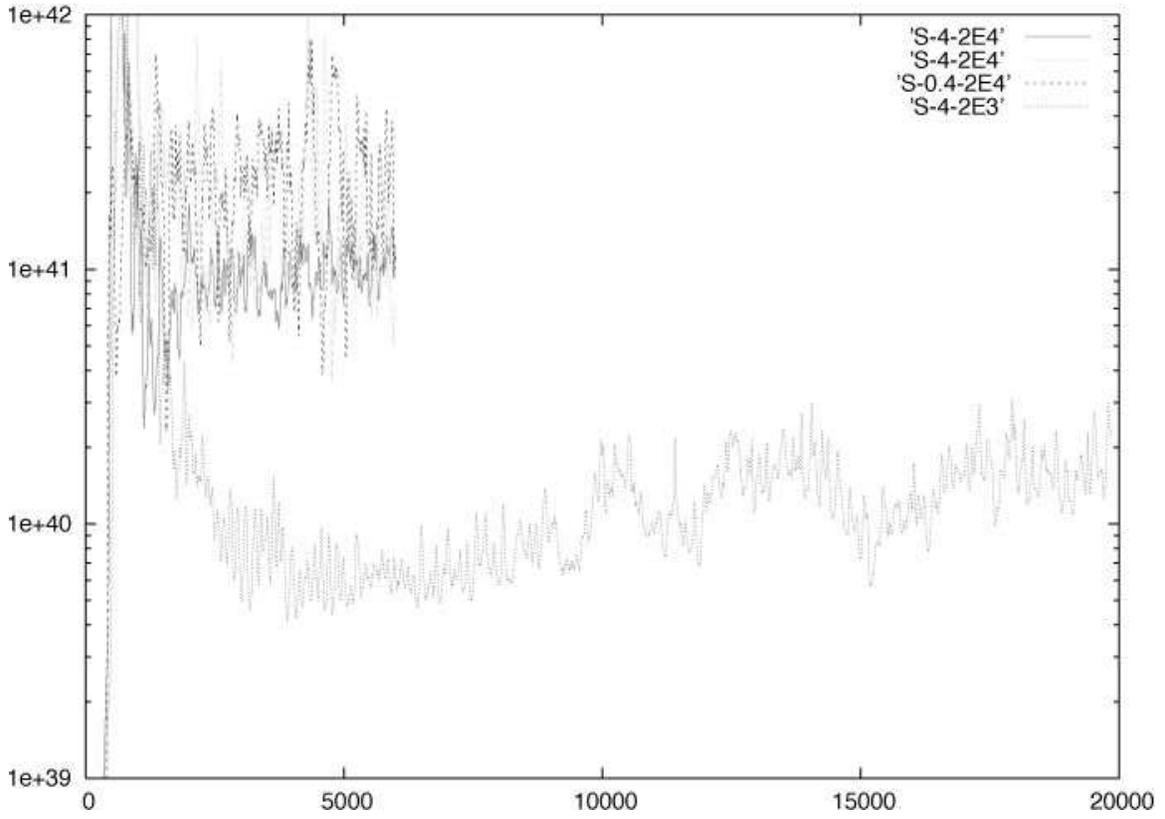}
\caption{Plot of Urca neutrino loss versus simulation time, of simulations 
S-4-2E4, S-1-2E4, S-0.4-2E4 and S-4-2E3. Simulations S-4-2E4, S-1-2E4, S-0.4-2E4 
had the nuclear rate multiplied by $2\times 10^4$ and simulation S-4-2E3
had the nuclear rate multiplied by $2\times 10^3$.
 }
\end{figure}

\clearpage

\begin{figure}
\figurenum{3}
\epsscale{0.4}
 \hskip +3truecm
 \includegraphics[width=2in]{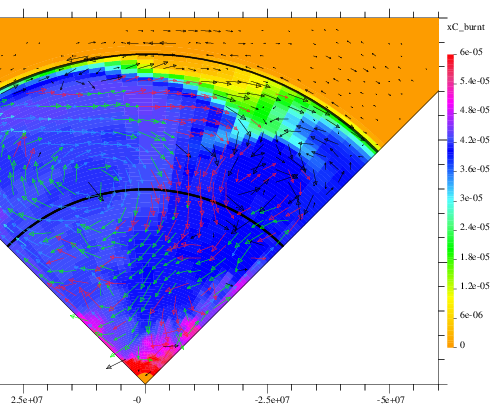}
 \hskip +3truecm 
 \includegraphics[width=2in]{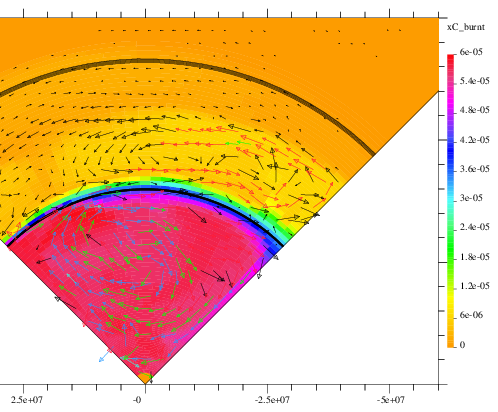} 
 \vskip +1truecm
 \hskip +3truecm
 \includegraphics[width=2in]{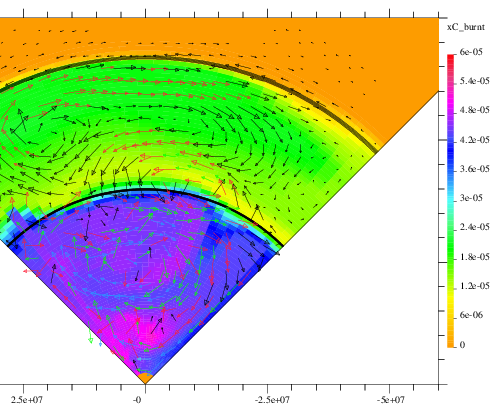} 
 \hskip +3truecm 
 \includegraphics[width=2in]{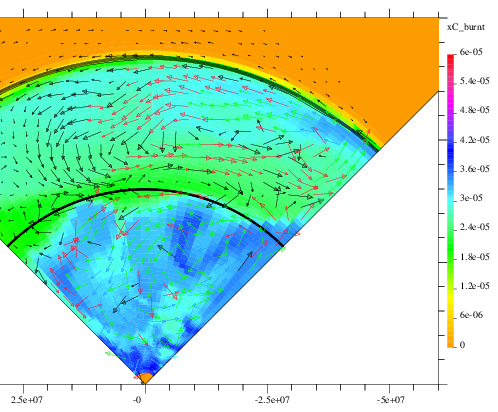} 
\caption{A) (upper left) Plot of the distribution of carbon-burning products, 
velocity vectors, the Urca shell (heavy black line around R = 400 Km) and the 
boundary separating the inner region of constant entropy from the outer 
isothermal region (thinner black line around 670 Km) for simulation S-4-2E4-0 
with no Urca process. The carbon-burning products are distributed more or less 
evenly in the whole inner region. B) (upper right) Same as A but for simulation
S-4-2E4 with the Urca process turned on .  The carbon-burning products are almost 
entirely confined to the inner zone (within the Urca-shell).  Strong convection is 
seen in the inner zone, and a separate convection region exists in the outer zone 
between the Urca shell and the boundary of the isothermal region, probably triggered 
by the overshoot from the inner zone. C) (lower left) same as B) but for simulation
S-1-2E4 with smaller abundance of Urca-active nuclei. The carbon-burning products 
are less confined to the inner zone (within the Urca-shell) than in B. The velocities 
near the Urca-shell are almost parallel to the shell.  D) (lower right) Same as B 
and C but for simulation S-0.4-2E4 with even smaller abundance of Urca-active nuclei. 
The carbon-burning products are less confined to the inner zone (within the Urca-shell) 
than in C.  Velocity vectors are normalized such that one horizontal tic mark represents
10 km s$^{-1}$ (journal black and white version) or black arrows: less than 1 
km s$^{-1}$;  red arrows: 1 - 2 km s$^{-1}$; green arrows: 2 - 4 km s$^{-1}$; 
blue arrows: 4 - 8 km s$^{-1}$ (on-line color version). 
}
\end{figure}

\clearpage

\begin{figure}
\figurenum{4}
\epsscale{0.4}
 \hskip +3truecm
 \includegraphics[width=2in]{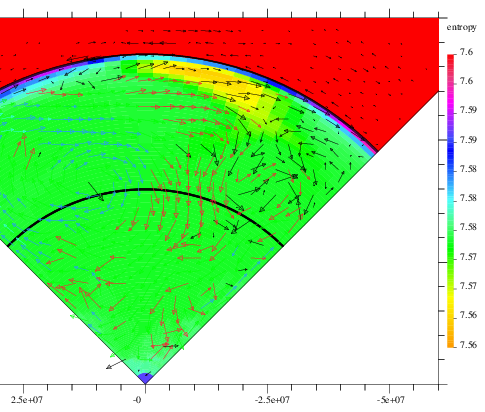}
 \hskip +3truecm 
 \includegraphics[width=2in]{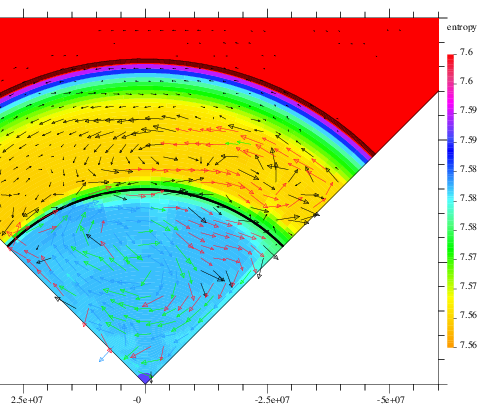} 
\caption{A) (left panel) Plot of entropy, velocity vectors, Urca shell as in Figure 3 
at 10,000 seconds for simulation S-4-2E4-0 with no Urca effect. B) (right panel)
same as A, but with Urca effect and time = 6000 sec for simulation S-4-2E4.  
 }
\end{figure}

\clearpage

\begin{figure}
\figurenum{5}
\epsscale{0.4}
 \hskip +3truecm
 \includegraphics[width=2in]{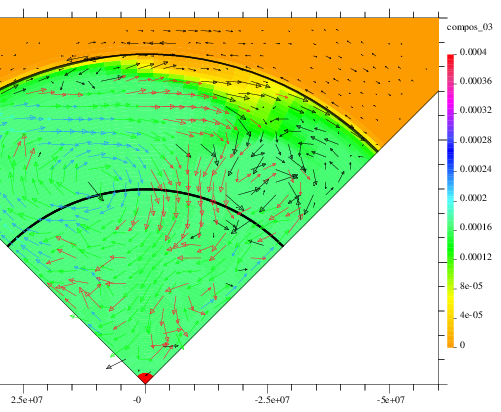}
 \hskip +3truecm 
 \includegraphics[width=2in]{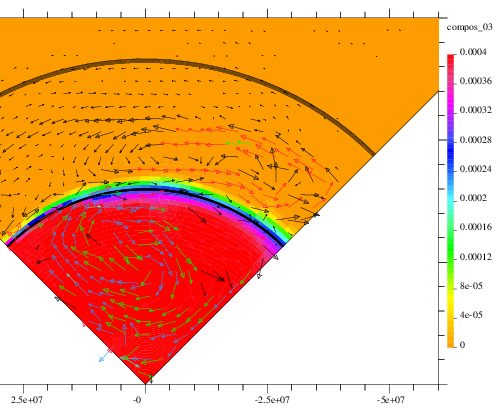} 
\caption{A) (left panel) 
Plot of $Ne^{23}$/$Na^{23}$ abundance, velocity vectors, Urca shell as in Figure 3
at 10,000 seconds for simulation S-4-2E4-0 with no Urca effect. B) (right panel)
same as A, but with Urca effect and time = 6000 sec for simulation S-4-2E4.
 }
\end{figure}

\clearpage

\begin{figure}
\figurenum{6}
\epsscale{0.4}
 \hskip 6truecm
\includegraphics[width=4in]{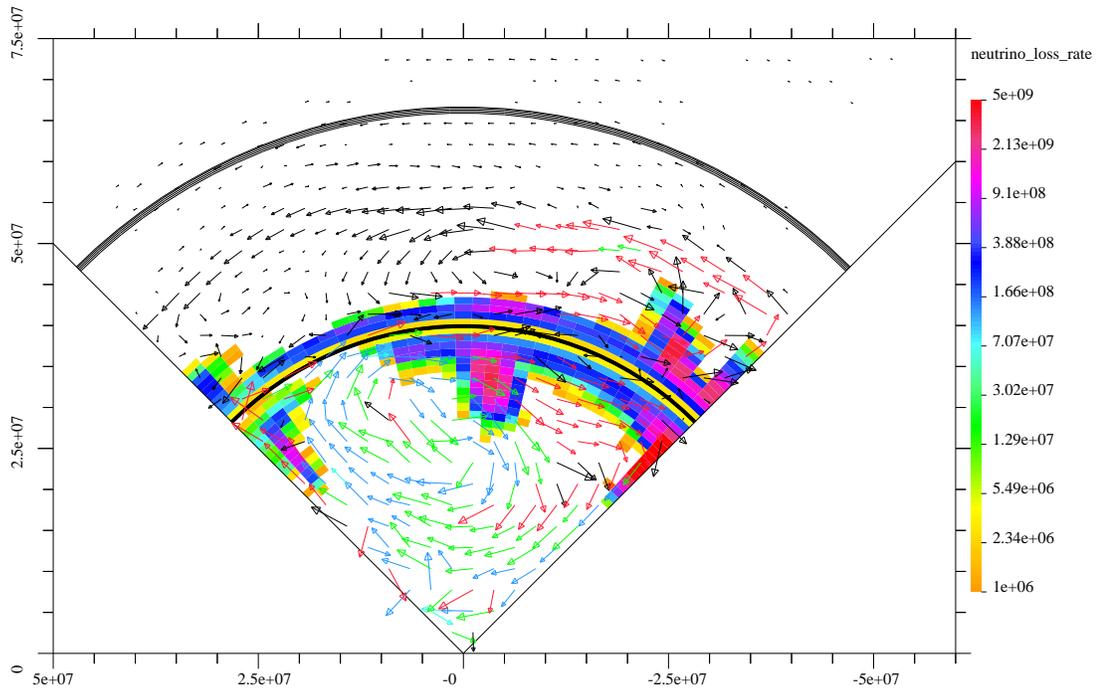}
\caption{Plot of Urca neutrino loss for simulation S-4-2E4. The whole Urca activity 
occurs near the Urca-shell. Areas adjacent to the left and right boundaries are 
subject to numerical artificats and should be ignored. Velocity vectors are represented 
as in Figure 3.
 }
\end{figure}

\end{document}